\documentclass{ws-procs975x65}
\begin{document}
\title{\uppercase{Massive gravity as a limit of bimetric gravity}}
\author{\uppercase{Prado Mart\'{i}n-Moruno}, \uppercase{Valentina Baccetti}, and \uppercase{Matt Visser}}
\address{School of Mathematics, Statistics, and Operations Research\\
Victoria University of Wellington \\
PO Box 600, Wellington 6140,
New Zealand}

\begin{abstract}
Massive gravity may be viewed as a suitable limit of bimetric gravity. 
The limiting procedure can lead to an interesting interplay between the ``background'' and ``foreground'' metrics 
in a cosmological context. The fact that in bimetric theories one always has two sets of metric equations of motion 
continues to have an effect even in the massive gravity limit. Thus, solutions of bimetric gravity in the limit of
vanishing kinetic term are also solutions of massive gravity, but the contrary statement is not necessarily true.
\end{abstract}


\bodymatter\bigskip

Massive gravity is the theory based on the consideration of a non-vanishing mass for the graviton.
As is well-known, to construct such a mass term in the action one needs an additional two-tensor, $f_{\mu\nu}$. 
Therefore, the dynamics of the ``foreground'' metric, $g_{\mu\nu}$, depends on a ``background'' metric, 
$f_{\mu\nu}$, which is specified by the definition of the theory.
On the other hand, in bimetric gravity the consideration of a spin-2 $f$-particle interacting with the metric $g_{\mu\nu}$
is based on a dynamical $f_{\mu\nu}$. That is, the new metric also has a kinetic term.
So one has two mutually interacting metrics with the same status.

Due to recent developments in the field\cite{deRham:2010ik, deRham:2010kj,Hassan:2011vm,Hassan:2011hr, Hassan:2011tf, Hassan:2011ea}
we now know that a mass term which is a particular function
of the square-root $\sqrt{g^{-1}f}$ is free of the the  Boulware--Deser 
ghost~\cite{Boulware:1973my}. 
Moreover, taking an interaction term of the same form in bimetric gravity leads also to
a ghost-free theory \cite{Hassan:2011zd}.
Considering
$g^{\mu\nu}=\eta^{AB}\,e^{\mu}{}_A\,e^{\nu}{}_B$ and $f_{\mu\nu}=\eta_{AB}\,w_{\mu}{}^A\,w_{\nu}{}^B$,
and assuming the condition\cite{Volkov:2011an} $e^{\mu}{}_A\;w_{B\mu}=e^{\mu}{}_B\;w_{A\mu}$,
one can write the square root as
$ \gamma^{\mu}{}_{\nu}=e^{\mu}{}_A\,w^A{}_{\nu}$.
Thus, the action of the ghost-free massive gravity theory can be expressed as \cite{Baccetti:2012bk}
\begin{equation}\label{actionmg}
S_{MG}=-\frac{1}{16\pi G}\int d^4x\sqrt{-g} \left\{R(g)+2\,\Lambda-2\,m^2 \, L_{{\rm int}}(\gamma)\right\}+S_{({\rm m})},
\end{equation}
where $S_{({\rm m})}$ describes the matter fields coupled to $g_{\mu\nu}$ and
\begin{equation}\label{int}
 L_{{\rm int}}=e_2(K)-c_3\,e_3(K)-c_4\,e_4(K),\quad
{\rm with}\quad
 K^{\mu}{}_{\nu}=\delta^{\mu}{}_{\nu}-\gamma^{\mu}{}_{\nu},
\end{equation}
where the $e_i$ are the elementary symmetric polynomials, given by
\begin{eqnarray}
&&
e_2(K)=\frac{1}{2}\left([K]^2-[K^2]\right);
\qquad 
e_3(K)=\frac{1}{6}\left([K]^3-3[K][K^2]+2[K^3]\right);\nonumber\\
&&
e_4(K)=\frac{1}{24}\left([K]^4-6[K^2][K]^2+3[K^2]^2+8[K][K^3]-6[K^4]\right);
\end{eqnarray}
where $[K]=K^\mu{}_\mu$. Therefore, the equations of motion for massive gravity are
\begin{equation}\label{motion1}
 G^{\mu}{}_{\nu}-\Lambda\,\delta^{\mu}{}_{\nu}=m^2\,T^{({\rm eff})\mu}{}_{\nu}+8\pi G \;T^{({\rm m})\mu}{}_{\nu} ,
\end{equation}
with $T^{({\rm m})\mu}{}_{\nu}$ denoting the usual stress energy tensor associated with $S_{({\rm m})}$,
\begin{equation}\label{T}
 T^{({\rm eff})\mu}{}_{\nu}=\tau^{\mu}{}_{\nu}-\delta^{\mu}{}_{\nu}\,L_{{\rm int}},
 \quad {\rm and} \quad
 \tau^{\mu}{}_{\nu}=\gamma^\mu{}_\rho\,\frac{\partial L_{{\rm int}}}{\partial \gamma^{\nu}{}_{\rho}}
 =e^{\mu}{}_B\,\frac{\partial L_{{\rm int}}}{\partial e^{\nu}{}_{B}}.
\end{equation}
The interaction term~(\ref{int}), therefore, leads to
\begin{eqnarray}\label{tau}
\tau=([\gamma]-3)\gamma-\gamma^2+c_3\left(e_2(K)\gamma-e_1(K)\gamma\cdot K+\gamma\cdot K^2\right)
 \nonumber\\
+ c_4\left(e_3(K)\gamma-e_2(K)\gamma\cdot K+e_1(K)\gamma\cdot K^2-\gamma\cdot K^3\right).
\end{eqnarray}
One must also take into account the Bianchi inspired constraint:
$\nabla_{\mu}T^{({\rm eff})\mu}{}_{\nu}=0$.
As in bimetric gravity both metrics are dynamical quantities, the action is
\begin{equation}\label{actionbg}
S_{BG}=S_{MG}-\frac{\kappa}{16\pi G}\int d^4x\sqrt{-f} \left\{\mathcal{R} (f)  +2\, \bar\Lambda \right\} +\epsilon\,\bar S_{({\rm m})} ,
\end{equation}
where $\bar S_{({\rm m})}$ is coupled to $f_{\mu\nu}$.
Thus in addition to Eq.~(\ref{motion1}) one has
\begin{equation}\label{motion2}
 \kappa\,\left(\mathcal{G}^{\mu}{}_{\nu}-\bar\Lambda\,\delta^\mu{}_\nu\right) =
m^2\, \mathcal{T}^{\mu}{}_{\nu} +\epsilon\,8\pi G \,\bar T^{({\rm m})\mu}{}_{\nu} ,
\quad{\rm with}\quad
\mathcal{T}^{\mu}{}_{\nu}=-\frac{\sqrt{-g}}{\sqrt{-f}}\;\tau^{\mu}{}_{\nu}.
\end{equation}
Furthermore, both Bianchi inspired constraints are equivalent.

It must be emphasized that whereas in massive gravity $f_{\mu\nu}$ is non-dynamical, 
in bimetric gravity there are two sets of equations of motion independently of the particular
limit that one considers. Thus, solutions of bimetric gravity are always more constrained than those of massive gravity.
If one wants to recover massive gravity as a limit of bimetric gravity, it should be noted that the consideration
of the limit of vanishing $\kappa$ and $\epsilon$ in Eqs.~(\ref{actionbg}) and (\ref{motion2}), leads to $S_{BG}=S_{MG}$ and
a constraint on the interaction term, respectively. While in contrast the limit of infinite $\kappa$ (and vanishing $\epsilon$)
in Eq.~(\ref{motion2})
leads to a constraint on the background metric. In the second case, $f_{\mu\nu}$ is no more externally specified, 
which is against the philosophy
of the theory.
Therefore, we conclude \cite{Baccetti:2012bk} that the limit 
of vanishing $\kappa$ and $\epsilon$, that is of vanishing kinetic term, must be taken in bimetric gravity to recover massive gravity.
Taking this limit in Eq.~(\ref{motion2}) implies $\tau^{\mu}{}_{\nu}=0$, which together with Eq.~(\ref{T}) leads
to $ \partial_{\lambda}L_\mathrm{int}=0$. Thus, the Bianchi inspired constraint can be written as $\partial_{\lambda}L_\mathrm{int}=0$,
which implies a modification equivalent to a cosmological constant, $T^{({\rm eff})\mu}{}_{\nu}=-\Lambda_{{\rm eff}}\delta^{\mu}{}_{\nu}$.
Thus, {\em the solutions of bimetric gravity in the limit of vanishing kinetic term are solutions of massive gravity in which
the modification with respect to general relativity is equivalent to a cosmological constant}. 
Therefore, it is of special
interest to consider the cosmological solutions.

For our present purposes, the metrics in a spherically symmetric situation can be written as \cite{Baccetti:2012bk}
\begin{equation}\label{metricg}
 g_{\mu\nu}\,dx^{\mu}\,dx^{\nu}=S^2\,dt^2-N^2\,dr^2-R^2\,d\Omega_{(2)}^2,
\end{equation}
and
\begin{equation}\label{metricf}
 f_{\mu\nu}\,dx^{\mu}\,dx^{\nu}=A^2 \,dt^2-B^2 \,dr^2-U^2\,d\Omega_{(2)}^2,
\end{equation}
where all the metric coefficients are functions of $t$ and $r$.
As has been shown in Ref.~\refcite{Baccetti:2012bk} the solutions of the system given by 
$\tau^{\mu}{}_{\nu}=0$ and $\partial_{\lambda}L_\mathrm{int}=0$ are:
\vspace*{-6pt}
\begin{itemlist}
\item {\em Conformally related metrics}: 
\begin{equation}
f_{\mu\nu}\,dx^{\mu}\,dx^{\nu}=D^2 \,g_{\mu\nu}\,dx^{\mu}\,dx^{\nu}
\quad with \quad
D = 1 + {3 c_3\over2 c_4} \pm \sqrt{ \left(1+ {3 c_3\over2 c_4}\right)^2- 1}. 
\end{equation}
Here $\Lambda_\mathrm{eff}>0$ if $c_3(1-D)>3$. 
Thus we can describe a universe which is accelerating,
with the physical metric having the same symmetry as the background.
\item {\em Conformally related sections}: 
\begin{equation}
f_{\mu\nu}\,dx^{\mu}\,dx^{\nu}=\bar{D}^2\,S^2(r,t)\,dt^2-\bar{D}^2\,N^2(r,t)\,dr^2-D^2\,R^2(r,t)\,d\Omega_{(2)}^2,
\end{equation}
with $D$ and $\bar D$ such that both $\bar{D}=\frac{-D+c_3(D-1)+2}{c_3(D-1)+1}$ and
$c_4=\frac{-1-c_3+c_3^2(D-1)^2-c_3D}{(D-1)^2}$. This solution has $\Lambda_\mathrm{eff}>0$ if $c_3(D-1)>-1$.
\end{itemlist}
\vspace*{-6pt}
\noindent  For massive models with $c_4=-3/4\,c_3^2$ there are two additional solutions \cite{Baccetti:2012bk}
with $\Lambda_{{\rm eff}}<0$. However, they have particular interest because
they allow us to consider different cosmological metrics related by some unconstrained arbitrary functions. In particular, these
solutions are:
\vspace*{-6pt}
\begin{itemlist}
 \item  {\em Independent $f_{rr}$}: We can obtain FLRW solutions if the theory specifies
 \begin{equation}
  f_{\mu\nu}\,dx^{\mu}\,dx^{\nu}=D^2\,a^2(t)\,dt^2-B^2(r,t)\,dr^2-D^2\,a^2(t)\,r^2\,d\Omega_{(2)}^2,
 \end{equation}
 where $B(r,t)$ is an arbitrary function which can be a function of $a(t)$ or not, and $a(t)$ is the scale factor fulfilling
 the Friedmann equation of the physical space. On the other hand, if the theory is formulated with a FLRW background, 
 then we can obtain isotropic or anisotropic cosmologies given by
 \begin{equation}
 g_{\mu\nu}\,dx^{\mu}\,dx^{\nu}=a^2(t)/D^2\,dt^2-N^2(r,t)\,dr^2-a^2(t)\,r^2/D^2\,d\Omega_{(2)}^2.
 \end{equation}
 \item  {\em Independent $f_{rr}$}: Similar behavior as in the previous case, with the unconstrained function now in the temporal part.
\end{itemlist}

Finally, it must be emphasized that there are, of course, more cosmological solutions of massive gravity. The solutions described in
this contribution
are only the cosmological solutions of massive gravity continuous in the parameter space of the theory, if one insists on the desirability of this theory coming from a particular limit of bimetric gravity.

\section*{Acknowledgments}
PMM acknowledges financial support from the Spanish Ministry of
Education through a FECYT grant, via
the postdoctoral mobility contract EX2010-0854.


\end{document}